%% file: main.tex
\documentclass{article}

\usepackage[margin=1in]{geometry}
\usepackage[english]{babel}
\usepackage[utf8]{inputenc}
\usepackage{csquotes}

\usepackage{subfiles}
\usepackage{graphicx}
\usepackage{placeins}
\usepackage{authblk}

\usepackage{hyperref}

\usepackage[style=numeric-comp,sorting=none,natbib=true]{biblatex}
\title{The Radicalization Pipeline:\\
Characterizing and Predicting Online Misogyny}

\date{}

\author[1]{Anna Bertani}
\author[2]{Nathan Ratkiewicz}
\author[2]{Alessandro Flammini}
\author[2]{Filippo Menczer}
\affil[1]{Complex Human Behaviour Lab, Fondazione Bruno Kessler, Trento, Italy}
\affil[2]{Observatory on Social Media, Indiana University, Bloomington, USA}

\begin{document}

\maketitle
    
\begin{abstract}
Social media platforms are known to shape and polarize user opinions and, increasingly, facilitate pathways toward ideological extremism. Among these trends, online misogynistic communities associated with the incel (involuntary celibate) ideology have received growing scholarly attention. Previous work has primarily focused on ideological narratives and linguistic content within incel communities themselves. However, much less is known about how user behaviors evolve over time, particularly before and after their engagement with incel spaces, and whether early signals can anticipate such transitions.
In this study, we conduct a longitudinal analysis of Reddit users who participated in the \emph{r/incels} community on Reddit. 
Our results show that engagement with the incel community is associated with a substantial increase in toxicity in both posts and comments, while hate  increases particularly in posts after engagement. These changes extend beyond explicitly ideological communities: users who engage with the \emph{r/incels} subreddit display elevated toxicity and hate speech even in subreddits related to leisure, entertainment, and everyday interests. In parallel, user activity shifts away from recreational and support-oriented communities toward subreddits focused on social issues, indicating a broader reorientation of online participation.
Finally, we show that future incel users can be identified with high accuracy prior to engagement using supervised machine learning models trained on linguistic, contextual, and interaction-based features. These findings suggest that online misogynistic radicalization follows detectable behavioral trajectories and diffuses across the broader social media ecosystem.
\end{abstract}

\section*{Introduction}

\subfile{introduction}

\section*{Methods}
    \label{sec:methods}

\subfile{methods}

\section*{Results}

\subfile{results}

\FloatBarrier

\section*{Discussion}

\subfile{discussion}

\section*{Acknowledgments}

We are grateful to Jason Baumgartner for the Pushshift Reddit dataset.  
AB gratefully acknowledges the financial support of the European Union's Erasmus+ Traineeship Programme for her research mobility at Indiana
University, and thanks Riccardo Gallotti for his support.
This research was supported in part by the Knight Foundation and the Luddy School of Informatics, Computing, and Engineering at Indiana University. 

\printbibliography
    

\renewcommand\thefigure{S\arabic{figure}}
\setcounter{figure}{0}
\renewcommand\thetable{S\arabic{table}}
\setcounter{table}{0}


\end{document}

%% file: introduction.tex
Social media platforms serve not only as channels for communication, but also as digital spaces in which users actively construct identities, negotiate social norms, and interpret their social realities \cite{boyd2010social, papacharissi2016affective}. These processes are also shaped by the affordances of specific platforms, such as hashtags on Twitter/X or the community-based architecture of Reddit, which encourage users to connect with like-minded others and cluster around shared interests \cite{williams2022identity, evans2017explicating}. While the creation of these groups can facilitate social support and community belonging, the same sociotechnical mechanisms that connect users around shared experiences also restrict their informational environments, isolating them from other perspectives. In this scenario, users are increasingly exposed to homogeneous viewpoints, reinforcing polarization and, in some cases, accelerating pathways toward radicalization \cite{del2016spreading, cinelli2021echo, ribeiro2020auditing}.

A growing body of research has examined how digitally networked spaces can serve as incubators for extremist ideologies by enabling sustained interaction, identity reinforcement, and the normalization of radical beliefs. Within this literature, particular attention has been paid to the misogynistic subcultures collectively referred to as the ``Manosphere'' \cite{ging2019alphas, marwick2018drinking}. The Manosphere is not a single unified movement, but rather a series of online communities connected by antifeminist sentiment and a shared perception that masculinity is under threat. It includes groups such as Men’s Rights Activists (MRAs), Pick-Up Artists (PUAs), and Men Going Their Own Way (MGTOW), which interact through overlapping discourse and membership \cite{gotell2016sexual, vanvalkenburgh2021digesting, ribeiro2021evolution}.

Among these groups, the incel (involuntary celibate) community has emerged as the most extreme. The term ``incel'' was originally coined in 1997 to describe an inclusive support network for people of all genders struggling with dating and intimacy \cite{incelterm}, known as ``Alana's Involuntary Celibacy Project.'' At this early stage, involuntary celibacy functioned primarily as a descriptive category rather than an ideological identity, and incel forums were used largely as peer-to-peer support spaces centered on shared vulnerability.

As online culture evolved in the early twenty-first century, however, these communities increasingly reflected the norms of participatory platforms such as 4chan and Reddit, where visibility and engagement were often driven by provocative and transgressive expression. Although incel forums were dispersed across multiple online spaces rather than confined to these platforms, their discourse gradually adopted similar stylistic conventions. Practices such as trolling and ``shitposting'' blurred the boundary between irony and sincerity, allowing expressions of frustration to escalate into increasingly hostile and exclusionary narratives \cite{hoffman2020assessing}. Over time, these dynamics contributed to the transformation of incel spaces from support-oriented communities into sites of ideological radicalization \cite{inceltermlater}. Indeed, the incel ideology promotes a form of biological determinism in which physical appearance is seen as genetically fixed and treated as the primary determinant of social and romantic worth. Those who fail to meet rigid beauty standards are considered irrevocably excluded from intimacy and social success \cite{glace2021taking, maxwell2020short, hoffman2020assessing}.

Scholars have described ``being incel'' in multiple dimensions: as a social identity (self-identifying as incel), as an ideology (adopting misogynistic beliefs consistent with incel views), and as group membership (belonging to incel communities) \cite{maryn2024identifying}. 
At the same time, these spaces often function as informal peer-support networks for individuals experiencing significant psychological distress. Empirical studies suggest that many users are drawn to incel communities due to depression, social anxiety, autism spectrum traits, and chronic isolation \cite{costello2024mating, moskalenko2022predictors}. Yet, rather than alleviating distress, community dynamics frequently encourage co-rumination \cite{fontanesi2024does, costello2022levels}, whereby users repeatedly externalize blame. Over time, personal vulnerability is reframed as collective injustice, increasing hostility toward women and society at large \cite{daly2022think, costello2022levels}. This process not only intensifies mental health struggles but also serves as a cognitive bridge to ideological radicalization \cite{speckhard2021involuntary}, contributing to the amplification of hateful discourse and the normalization of violence against women \cite{jaki2019online}.  

These dynamics have important implications for how individuals come to enter and participate  within incel spaces. 
As incel communities became increasingly associated with misogynistic extremism and offline violence, platforms decided to intervene and, in 2017, Reddit banned its largest incel forum, \emph{r/incels}. Although such interventions aim to curb hate speech and reduce harm, research on platform governance suggests their effects are mixed. Studies of deplatforming indicate that while bans disrupt established communities, they often incentivize users to migrate to alternative subreddits or less regulated platforms such as 4chan and dedicated forums, where extremist discourse may proliferate with fewer constraints \cite{chandrasekharan2017you, horta2021platform, ali2021understanding}. At the same time, recent evidence from the European Parliament suggests that these communities are not only displaced to fringe platforms, but also increasingly emerge within mainstream social media environments such as X, contributing to the broader normalization and diffusion of misogynistic narratives \cite{bentzen2026women}.

Despite an important body of research on incel ideology, discourse, and moderation strategies, an important methodological gap remains. Most existing studies rely on static analyses of content produced within incel communities. As a result, less is known about how user behaviors evolve over time, both before and after engagement, and which signals may anticipate entry into these spaces.

In this study, we examine the dynamics of engagement with incel communities through a longitudinal computational analysis of Reddit users who participated in the \emph{r/incels} subreddit. We address three research questions: (i)~how user levels of toxicity and hate speech change following engagement with the incel community compared to their pre-engagement activity; (ii)~whether users who eventually join incel communities are distinguishable from other Reddit users before their first direct participation, based on linguistic, contextual, and behavioral features; and (iii)~to what extent patterns of social interaction, particularly participation in overlapping discussion threads, differentiate future incel users from non-incel users, indicating the role of social influence in pathways of engagement. Our analyses allow us to identify early linguistic and behavioral features of radicalization, contributing to broader debates on how platform architectures intersect with individual vulnerability to shape extremist identities online.

%% file: methods.tex
\subsection*{Overview of the Dataset}

We collected Reddit data from the \emph{r/incels} subreddit using the Pushshift API \cite{baumgartner2020pushshift}. 
The original dataset includes 54,328 posts and 732,377 comments spanning the period from January 4, 2014 to November 3, 2017, before the subreddit was banned on November 7, 2017, for violating the Reddit policy on content that calls for violence or physical harm \cite{news-incel-suspension}. 

In our analysis, we focus on the most active users in the incel community. We define a user as active if, within \emph{r/incels}, they meet three criteria: (i)~they commented in at least five distinct months, not necessarily consecutive, during the period covered by our data; (ii)~they posted at least four comments in each month in which they were active; and (iii)~they made at least one post.
While such thresholds are somewhat arbitrary, they help us exclude users whose participation in the community was only incidental or occasional. 

We traced the online behavior of this subset of active incel users (henceforth ``incel users'' or ``incels'') across other communities both before and after their first activity (post or comment) in the \emph{r/incels} subreddit. This resulted in a dataset comprising 27,200 posts and 519,001 comments in 5,009 unique subreddits by 195 active incel users.

To investigate differences between active incel users and those who did not engage in incel-related discussions, we also extracted comments by users who never joined or participated in such community, but who contributed to other subreddits in which users were involved prior to joining the \emph{r/incels} subreddit.  This second dataset consists of 605,922 comments posted by 293,098 users (henceforth ``non-incel users'') across 1,595 unique threads in 281 unique subreddits.

\subsection*{Subreddit Categories}

To analyze the thematic structure of user participation across Reddit, we categorized each subreddit into broader topic domains. We employed a zero-shot classification approach using ChatGPT (GPT-3.5) to automate this task. 

In an initial phase, the model was prompted with only the name of the subreddit and asked to infer a suitable thematic label, without being given any predefined list of categories or labeled examples. 
After collecting these outputs, we identified the 10 most frequently occurring and semantically consistent categories. 

In a second phase, we queried ChatGPT again for each subreddit, this time instructing it to classify it into one of these 10 already defined categories.
Each subreddit was therefore classified in one of the following categories: Entertainment, Gaming, Lifestyle, Mental Health \& Wellness, Parenting \& Family, Personal Development, Relationship, Social Issues, Technology \& Science, Work \& Employment. 

\subsection*{Toxicity and Hate Speech}

We assigned toxicity scores using the Perspective API \cite{lees2022new}, a machine learning model that assigns a piece of text a score between zero and one, where one is most toxic. The API defines as toxic a comment that is ``very hateful, aggressive, disrespectful [\dots] or otherwise very likely to make a user leave a discussion or give up on sharing their perspective.''
The Perspective API also offers other scores including severe toxicity, identity attack, insult, profanity, threat, and sexually explicit. These scores were computed for each comment and post in our dataset and used as both individual and contextual features in our analysis.

To detect explicit hate speech, we used HateBERT, a language model based on BERT-base and further pre-trained on over one million posts from Reddit communities that were banned for violating platform-wide hate speech policies \cite{caselli2020hatebert}. This model has been shown to outperform general-purpose language models on hate speech detection tasks by capturing domain-specific language patterns prevalent in toxic and hateful online spaces. Each comment in our dataset was passed through HateBERT to determine whether it contained hate speech, resulting in a binary label.

\subsection*{Lexical Feature Analysis}

Previous work has shown that the incel community has an extensive cryptolect, or set of ingroup words that community members can use to identify themselves and that reflect the group's worldview. Klein et al. \cite{klein2024lexicon} developed a Lexicon that contains the words and building blocks of terminology common to incels, including terms such as \textit{toilet} (derogatory term for woman), \textit{rice} (prefix referencing East/Southeast Asians, e.g., ``ricecel''), \textit{manlet} (short man, used derogatorily), and \textit{foid} (female humanoid). 
We tracked the presence of these terms across posts and comments to quantify the adoption of incel-specific language.

\subsection*{Classifier Features}

We implemented a supervised machine learning framework to investigate whether it is possible to distinguish users who eventually join incel community from those who do not. 
Our classifier is based on user-level features extracted from posting behavior, interactions, and linguistic characteristics, with a binary target class variable indicating whether or not a user joined the incel community.

In particular, the features can be categorized into five types: (1)~toxicity and sentiment at the individual level; 
(2)~contextual toxicity and sentiment, computed as the average toxicity and sentiment of all comments preceding the target user's contribution within a thread; 
(3)~contextual user metrics, such as the number of unique users who commented before the target user; 
(4)~user activity and engagement metrics, including the number of likes received, number of comments posted, and number of unique subreddits in which the user participated; 
and (5)~linguistic features, such as the presence of lexicon commonly used by incels.

Given the class imbalance of our dataset (293,098 users who never joined the incel community and 195 who did), we employed a downsampling technique to balance the dataset, creating 10 different balanced datasets.
Prior to training, the features were rescaled using z-score normalization \cite{pedregosa2011scikit}.

We compared five machine learning models: Random Forest, Decision Tree, Logistic Regression, Support Vector Machine (SVM), and K-Nearest Neighbors (K-NN).
For each model, we trained and evaluated it on each of the 10 balanced datasets using 5-fold cross-validation. We report the mean precision, recall, F1-score, and AUC values averaged across all folds and datasets in Table ~\ref{tab:model_comparison}.

%% file: results.tex
\subsection*{Behavior Changes Following Incel Engagement}

To examine how incel users change before and after engaging with the incel community, we analyzed their behavior in terms of (i)~toxicity and hate speech, and (ii)~posting patterns, considering both the time of posts and their distribution across subreddits. 
We define pre-engagement as all comments and posts made by users before they became active in the incel community, and post-engagement as those made only after their first comment or post within the \emph{r/incels} subreddit.

\begin{figure}[htpb]
\centering
\includegraphics[width=\linewidth]{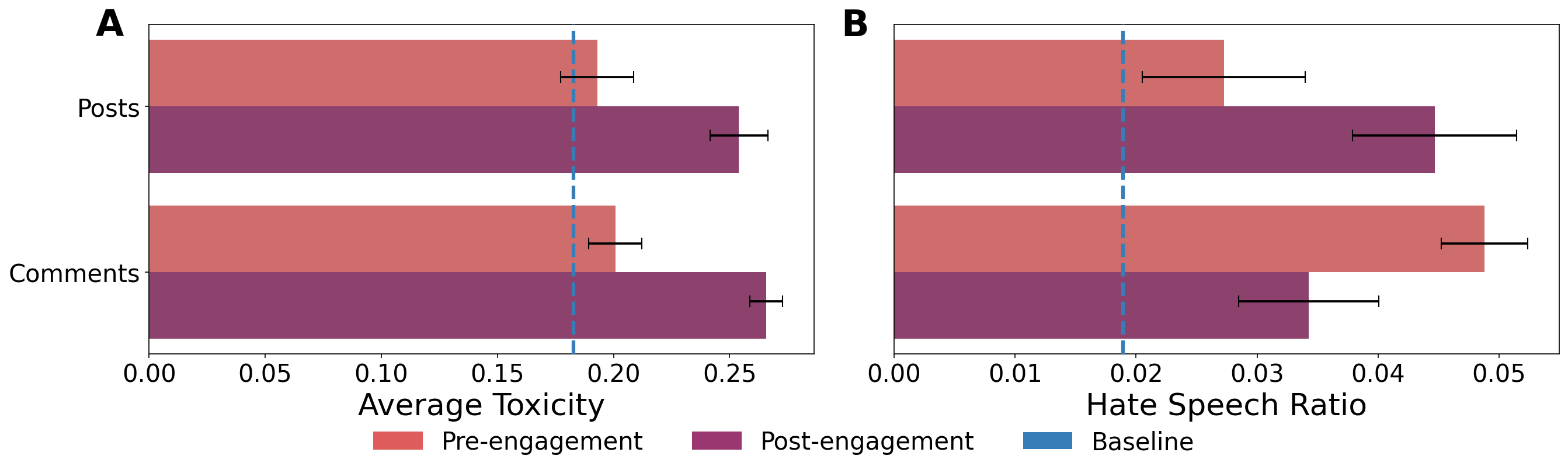}
  \caption{\textbf{Comparison of toxicity and hate speech levels in posts and comments before and after first engagement with the incel community.} (A)~Average toxicity scores, showing a consistent increase post-engagement relative to both pre-engagement activity and the baseline group of non-incel users (posts $p = 0.005$; comments $p < 0.001$). (B)~Proportion of content classified as hate speech, which increases after engagement for posts ($p = 0.04$) and decreases for comments ($p < 0.001$). All $p$-values are calculated using Mann--Whitney $U$ tests. The vertical dashed line shows the baselines of toxicity and hate speech for non-incel users.}
  \label{fig:toxicity_hate_general}
\end{figure}

Figure~\ref{fig:toxicity_hate_general}A summarizes changes in toxicity. Overall, users become significantly more toxic on average after engaging with the incel community. 
For comparison, the baseline group of non-incel users shows consistently lower toxicity, even compared to incel users in their pre-engagement phase.
Figure~\ref{fig:toxicity_hate_general}B shows the proportion of a user's posts or comments labeled as hate speech in each period.  
In posts, users express significantly more hate speech after engagement than before. 
In comments, however, the pattern reverses: users express significantly less hate speech after engagement. 
As with toxicity, the baseline remains much lower than both pre- and post-engagement levels.

\begin{figure}[h]
\centering
\includegraphics[width=\linewidth]{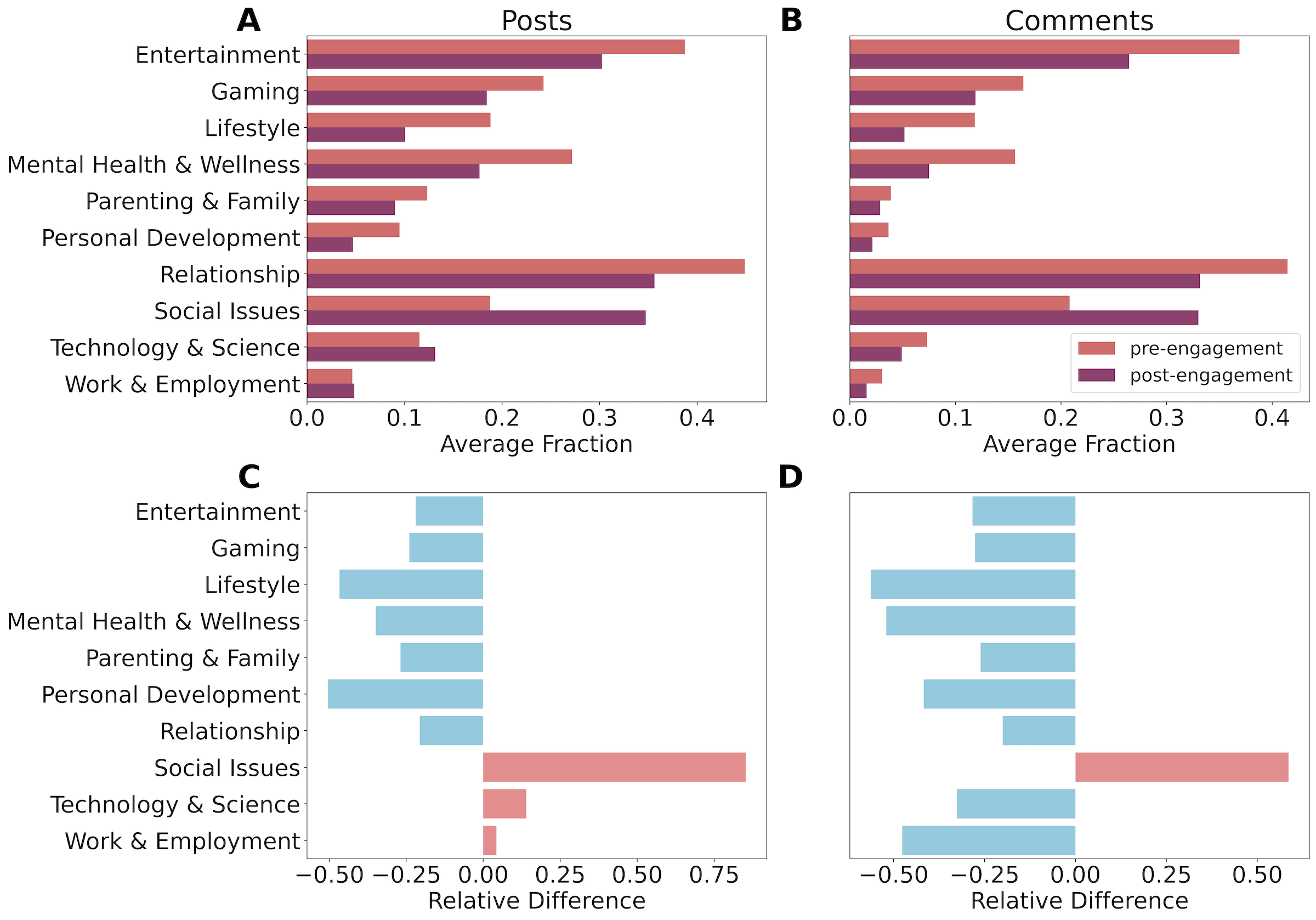}
  \caption{\textbf{Average distribution of user activity across subreddit categories before and after joining the incel community.} Panels show the mean fraction of (A)~posts and (B)~comments per category pre- and post-engagement, and the relative changes in (C)~posts and (D)~comments per category between pre- and post-engagement. We observe a pronounced shift toward subreddits about social issues and declining participation in most other topics.}
  \label{fig:fraction_posts_comments}
\end{figure}

Figure~\ref{fig:fraction_posts_comments} reports on how incel user posting patterns shifted across different subreddit categories. For each user and time period (pre vs. post), we first counted their total posts and comments and then computed the fraction falling into each content category. These fractions were then averaged across all users, leading to the mean distribution of activity across topics.
Before joining the incel community, users were primarily posting in subreddits focused on relationships, entertainment, mental health and wellness, gaming, and lifestyle, showing interest in leisure activities and participation in advice and/or support-oriented groups.
Participation was lower in subreddits related to social issues, which comprised the most radicalized communities. 
After engagement, incel users tended to post more in subreddits about relationships and social issues, followed by entertainment and gaming. 
Most categories tended to decrease after engagement, except for social issues, which increased by 85\%, as illustrated in Figure~\ref{fig:fraction_posts_comments}C. 
Looking at comments, users focused on relationships and entertainment, followed by social issues, gaming, and mental health and wellness before the transition to the \emph{r/incels} subreddit. 
Similarly to the posting behavior, after incel community engagement, users tended to comment most frequently in relationship and social issues channels, while interest for entertainment and gaming decreased. 
Figure~\ref{fig:fraction_posts_comments}D highlights how social issues were the only topics with increasing comment activity (by about 58\%) after engagement with the \emph{r/incels} subreddit. 

\begin{figure}[htpb]
\centering
\includegraphics[width=\linewidth]{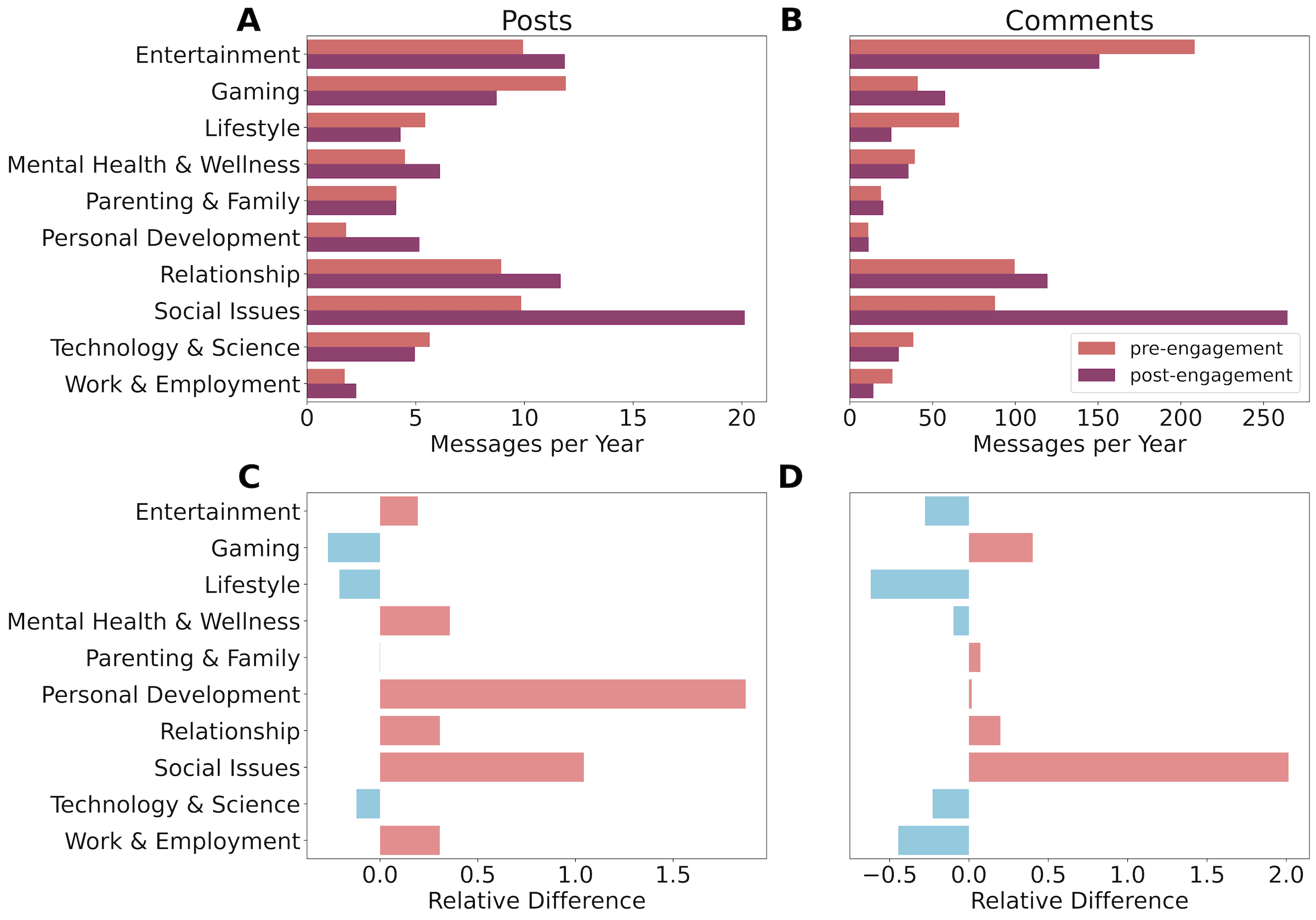}
  \caption{\textbf{Temporal evolution of user activity across subreddit categories before and after joining the incel community, normalized by time.} 
  Panels show the average annual number of (A)~posts and (B)~comments per category pre- and post-engagement, and the relative changes in (C)~posts and (D)~comments per category between pre- and post-engagement. We observe substantial growth in activity within social issues subreddits following engagement with the incel community, as well as an increase of posts (but not comments) about personal development.}
  \label{fig:timevariation_posts_comments}
\end{figure}

Shifting from the relative allocation of activity across topics to absolute activity levels, consider that the overall activity of a user depends on how long the user was active before and after joining the incel community. 
To analyze the temporal shift in absolute user activity while accounting for such differentials, let us analyze and compare the average activity of users per unit of time.
Figure \ref{fig:timevariation_posts_comments} shows the absolute values in terms of messages and comments posted per year as well as the relative differences between the pre- and post-engagement phases. We notice that users were initially more active in gaming and entertainment. After incel engagement, the most active category was social issues, with a relative increase of over 100\% in posts per year and over 200\% in comments per year. Annual posts in personal development reddits also increased by 187\% compared to an initially low activity, while comments did not increase. 
The category with most comments, entertainment, had a significant decrease in comment activity.  

\begin{figure}[htpb]
\centering
\includegraphics[width=\linewidth]{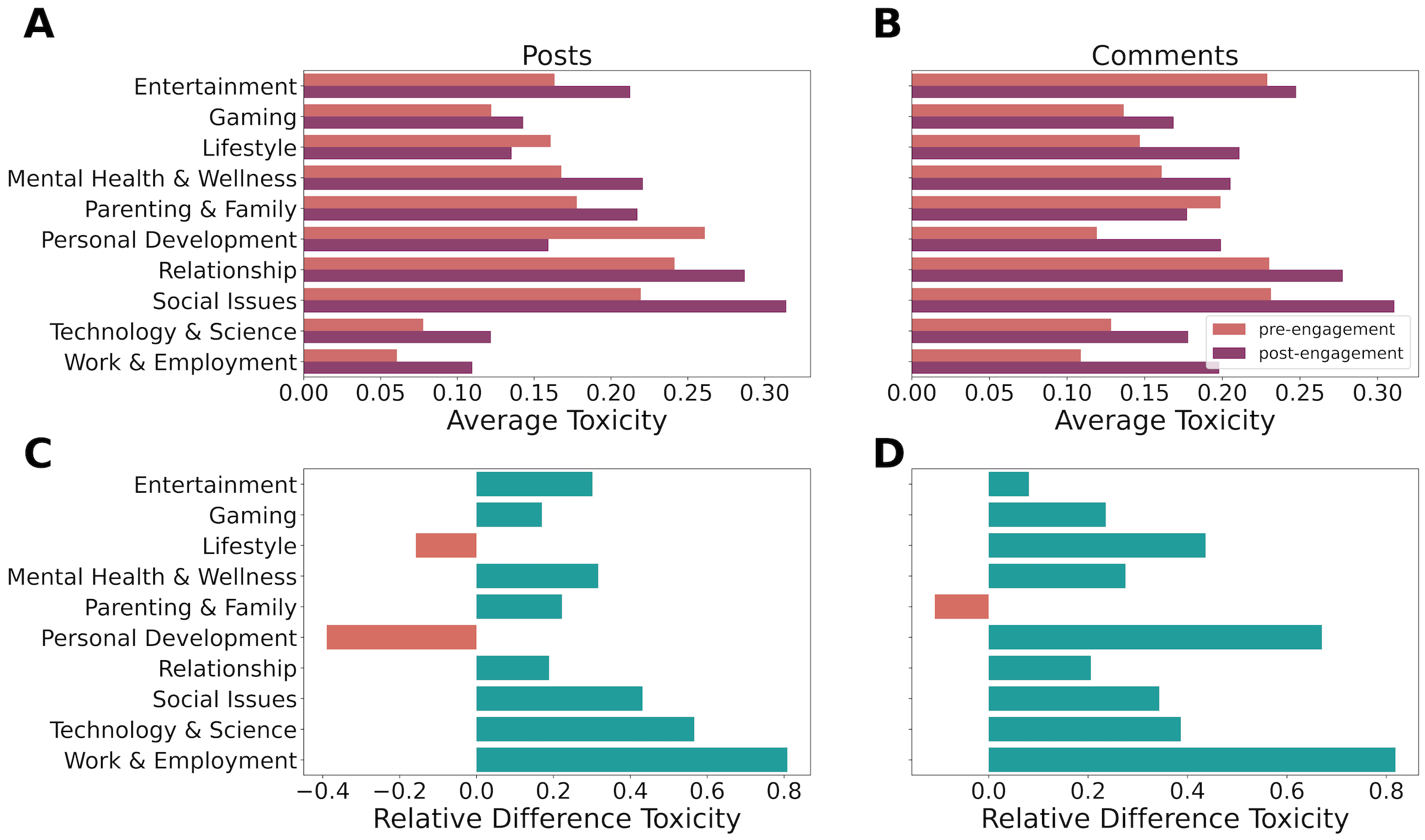}
  \caption{\textbf{Macro-averages of toxicity levels across subreddit categories before and after engagement with the incel community.} Panels report absolute toxicity values for (A)~posts and (B)~comments, as well as relative differences for (C)~posts and (D)~comments. We observe that toxicity increases after incel engagement across most categories, including those not explicitly political or ideological.}
  \label{fig:toxicity_subreddits}
\end{figure}

Beyond activity patterns, let us analyze the changes in the use of hate speech and toxicity after incel engagement across subreddit categories. For each of the 10 categories, we calculated the average toxicity per user, both pre- and post-engagement. Figure~\ref{fig:toxicity_subreddits}A,C reports the macro-averages of post toxicity and corresponding relative changes between pre- and post-engagement. Toxicity increases after incel engagement in most categories. The relative increase is highest in some categories with low pre-engagement toxicity, like work \& employment. The categories with the highest average post toxicity pre-engagement are personal development, relationship, and social issues. Among these, toxicity increases most prominently post-engagement in subreddits about social issues, while it decreases in personal development subreddits. 
Similarly for comments (Figure~\ref{fig:toxicity_subreddits}B,D), average toxicity increases post-engagement in most categories, especially for social issues in absolute terms and work \& employment in relative terms. 

\begin{figure}[htpb]
\centering
\includegraphics[width=\linewidth]{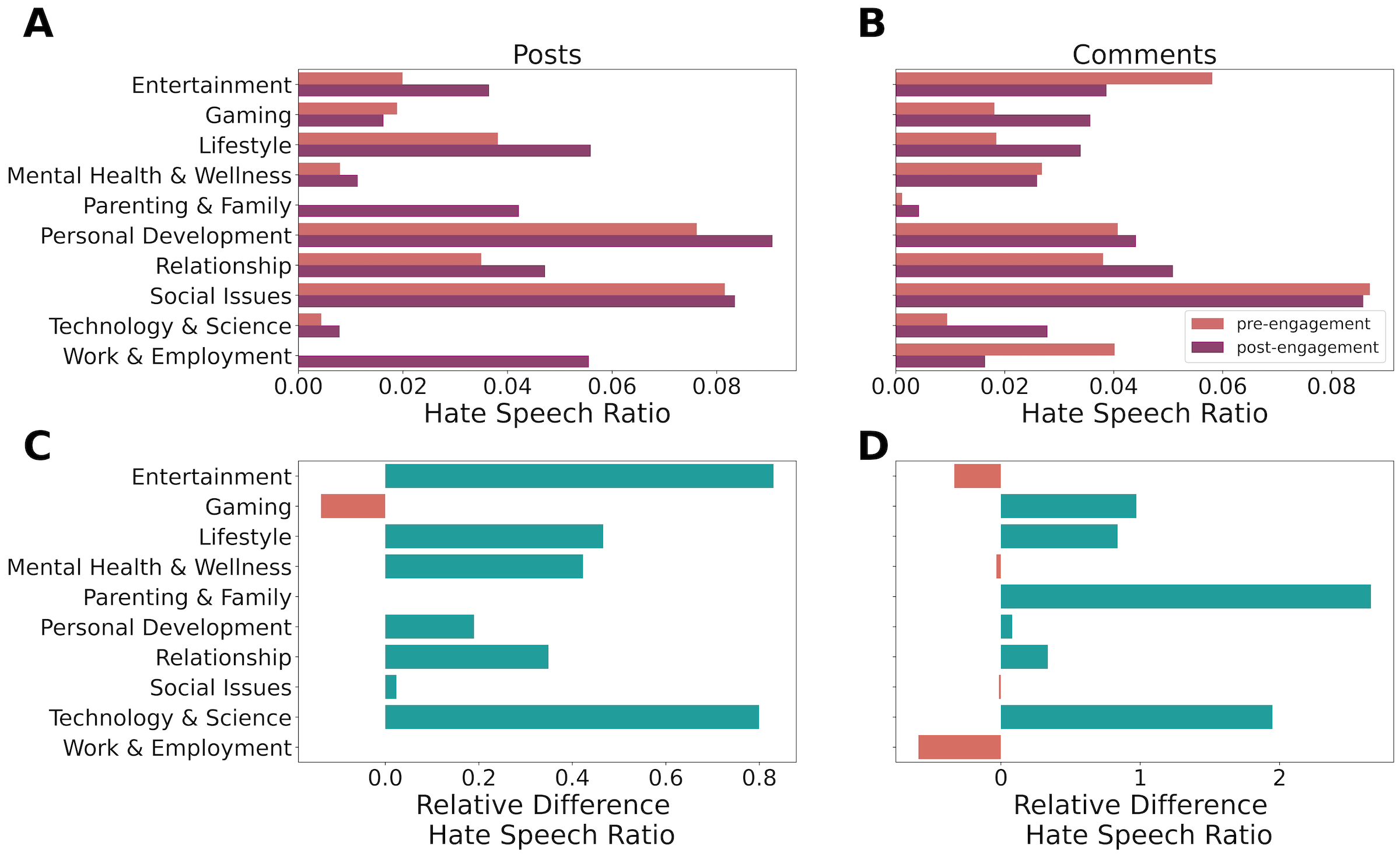}
  \caption{\textbf{Distribution of hate speech ratios across subreddit categories before and after incel engagement}. The ratios in absolute terms are reported for (A)~posts and (B)~comments, while the relative changes after engagement are reported for (C)~posts and (D)~comments, revealing an expansion of hate speech into a broad range of topical domains. Relative differences are undefined in categories with no pre-engagement hate speech.}
  \label{fig:hatespeech_ratio_categories}
\end{figure}

Figure \ref{fig:hatespeech_ratio_categories} reports on the analogous analysis for the ratio of hate speech posts and comments, pre- and post-engagement with the incel community, across subreddit categories. 
The categories with most hate speech posts before and after incel engagement are social issues and personal development. 
After engagement, hate speech posts become more prevalent in all categories except gaming. The largest relative increases are in categories where no hate speech was observed pre-engagement: work \& employment and parenting \& family. Even categories related to leisure activities, such as entertainment and technology \& science, show large relative increases in hate speech after engagement. 
Similar results are found when it comes to comments: the comment sections of subreddits about social issues are the primary channels where hate speech is spread. The largest growth in hate speech post-engagement is observed in parenting \& family and technology \& science. These increases are much larger than observed in posts --- hateful comments double or triple in these subreddits.

\subsection*{Prediction of Incel Engagement}

Can we predict whether a user will eventually join an incel community, based on their posting patterns? To explore this question, we evaluated several predictive machine learning models. Table~\ref{tab:model_comparison} reports the average F1 score, precision, recall, accuracy, and AUC for each model across 5-fold cross-validation repeated over the 10 datasets. Among all models, Random Forest achieved the highest predictive performance, suggesting that ensemble methods provide robust classification in this context.

\begin{table}[t]
\centering
\caption{\textbf{Comparison of supervised machine learning models for predicting future incel community membership.} We report precision, recall, F1 score, accuracy, and AUC averaged across repeated cross-validation. Highest metric values are shown in bold.}
\begin{tabular}{lccccc}
\hline
\textbf{Model} & \textbf{F1} & \textbf{Precision} & \textbf{Recall} & \textbf{Accuracy} & \textbf{AUC} \\
\hline
Decision Tree &  0.809 & 0.803 & 0.828 & 0.806& 0.850  \\
K-NN          &  0.867 & \textbf{0.914} & 0.831 &  0.874& 0.912 \\
Logistic Regression  &  0.843 & 0.875 & 0.827 & 0.847 & 0.924 \\
Random Forest &  \textbf{0.895} & 0.913 & \textbf{0.887} & \textbf{0.899} & \textbf{0.951} \\
SVM              &  0.869 & 0.892 & 0.856 & 0.873 & 0.926
 \\
\hline
\end{tabular}
\label{tab:model_comparison}
\end{table}

To assess the contribution of different types of features to predictive performance, we trained the Random Forest model using only one feature set at a time. As shown in Table~\ref{tab:feature_metrics}, all features are quite revealing. Contextual user metrics and linguistic features are among the most informative based on F1 scores. Toxicity and sentiment features, both individual and contextual, contribute most substantially based on AUC scores. User activity and engagement features are less predictive. 
These results suggest that a combination of contextual, linguistic, and toxicity-based features could be used in early interventions targeting users at risk of radicalization in misogynistic communities.

\begin{table}
\centering
\caption{\textbf{Feature analysis.} We report the prediction performance of Random Forest models trained on individual feature sets, illustrating the relative contribution of contextual interaction metrics, linguistic features, toxicity measures, and user activity indicators to classification accuracy. Highest metric values are shown in bold.}
\begin{tabular}{lcccc}
\hline
\textbf{Feature Set} & \textbf{Precision} & \textbf{Recall} & \textbf{F1} & \textbf{AUC} \\
\hline
Contextual User Metrics & 0.8218 & \textbf{0.8742} & \textbf{0.8469} & 0.8797 \\
Linguistic Features & 0.8216 & 0.8483 & 0.8319 & 0.8749 \\
Toxicity and Sentiment (Individual) & 0.8653 & 0.7967 & 0.8271 & 0.9040 \\
Toxicity and Sentiment (Context) & \textbf{0.8805} & 0.7850 & 0.8230 & \textbf{0.9171} \\
User Activity and Engagement & 0.7415 & 0.7608 & 0.7458 & 0.8499 \\
\hline
\end{tabular}
\label{tab:feature_metrics}
\end{table}

\subsection*{Social Influence}

\begin{figure}[t]
\centering
\includegraphics[width=0.6\linewidth]{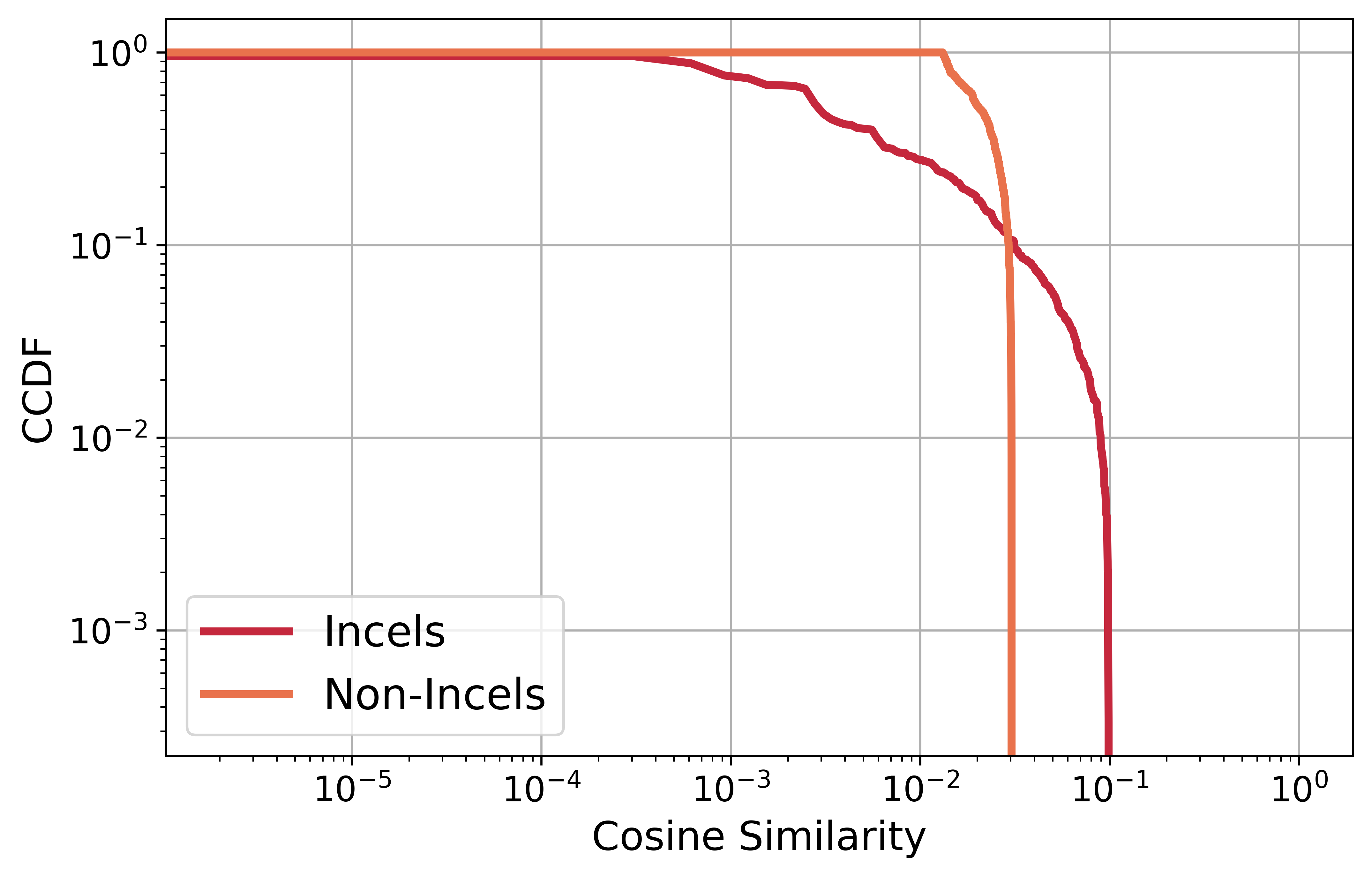}
  \caption{\textbf{Complementary cumulative distributions of cosine similarity scores measuring overlap in thread participation among non-incel and incel users during the pre-engagement phase}. The longer tail observed for incel users indicates greater behavioral similarity and shared interaction spaces, supporting the role of social influence in pathways toward community entry.}
  \label{fig:cosine_similarity}
\end{figure}

To investigate whether prior interactions in online discussion threads might be predictive of a user's transition to the incel community, we analyzed patterns of commenting behavior among incel users and users who never joined the incel community. In both cases, the analysis focused on threads where incel users were active prior to joining the r/incels subreddit. 
We first constructed a user-thread matrix, where each row corresponds to a user and each column corresponds to a thread, with counts of comments as values. To account for differences in activity across threads, we applied a TF-IDF transformation to the matrix, producing vector representations of user behavior.
Pairwise cosine similarities were then computed between incel users and, for computational feasibility, between non-incel users in randomly sampled groups of equal size. These similarities quantify the degree of participation in shared threads among users.
Figure~\ref{fig:cosine_similarity} shows the distributions of similarity for both groups of users. 
Despite the sparsity of user-thread interactions (most user pairs never commented in the same thread), the distributions differed significantly in their tails. A Mann–Whitney U test confirmed this difference 
($p < 0.001$), 
indicating that incel users are significantly more similar to each other in terms of their commenting patterns than non-incel users.
This analysis suggests that prior interactions and participation in the same discussion threads may provide a signal of social influence: users who eventually transition to the incel community tend to engage in more shared conversations compared to non-incel users.

%% file: discussion.tex
The findings of this study help us understand how online behaviors and conversation patterns evolve during the radicalization process toward misogynistic online communities such as \emph{r/incels}. By analyzing user activity before and after joining the \emph{r/incels} subreddit, we have demonstrated a behavioral and linguistic shift that reflects broader processes of ideological alignment and social reinforcement within extremist digital spaces.

A key finding is the significant increase in toxicity following user engagement with the incel community. Both comments and posts are characterized by higher average toxicity scores after joining, indicating that exposure to radicalized discourses not only normalizes but amplifies hostile communication patterns. Interestingly, while hateful posts increased considerably after incel engagement, hateful comments declined. This difference may suggest a strategic shift in expression: users might adopt more hateful discourse when creating original content, while engaging in subtler or coded forms of communication in discussion threads. The overall trend highlights how participating in extremist communities reshapes communicative norms and lowers the threshold for aggressive or demeaning language use.

The thematic analysis of subreddit participation further supports the notion of ideological drift. Prior to joining \emph{r/incels}, users were predominantly active in communities centered on entertainment, gaming, and relationships --- spaces typically characterized by social exchange or leisure. After joining, however, there was a clear shift toward ``social issues'' subreddits, with an 85\% increase in engagement relative to the pre-engagement phase. This suggests that incel users gradually reoriented their attention from personal or recreational interests toward socio-political arenas, often characterized by higher ideological polarization. 

Toxicity and hate speech were not uniformly distributed across categories. The post-engagement increase in toxicity within ``Social Issues,'' ``Relationships,'' and ``Mental Health \& Wellness'' subreddits suggests that radicalized discourse permeates even those communities traditionally oriented toward support and discussion. The expansion of hate speech into otherwise non-political categories like ``Entertainment'' and ``Technology \& Science'' further indicates the diffusion of extremist lexicons across the wider Reddit ecosystem. This diffusion effect represents a critical challenge for content moderation: radicalized narratives may propagate indirectly through mainstream spaces rather than remaining isolated within fringe communities.

Our machine learning results demonstrate that radicalization pathways are not random and can be predicted from behavioral and linguistic signals. 
Contextual user metrics and linguistic features, particularly those tied to incel-specific lexicon, are the strongest predictors of future membership. 
These findings imply that radicalization manifests not only through toxic content but also through subtle shifts in vocabulary. Early detection frameworks considering such features could thus serve as valuable tools for identifying at-risk users before full immersion in extremist subcultures occurs.

The social pattern analysis reveals another important result. The significantly higher cosine similarity in commenting behavior among future incel members suggests that radicalization is not purely individual, but also socially mediated. Users who eventually joined the \emph{r/incels} subreddit were more likely to interact within the same discussion threads, forming a web of exposure and reinforcement that may have facilitated entry into extremist spaces. 
This study highlights that online misogynistic radicalization is both predictable and dynamic, emerging from a series of individual and contextual characteristics as well as social interaction and exposure to normalized toxicity. Understanding this process points not only toward the mechanisms of digital extremism, but also toward data-driven interventions capable of mitigating its spread. 


Our results have several limitations. First of all, our reliance on Reddit data implies that our findings capture only a portion of radicalized content within misogynistic communities. Previous research has shown that incel-related discourse also occurs on other platforms, such as dedicated forums~\cite{yoder2023identity} and social media sites like YouTube~\cite{papadamou2021over}. Second, our analysis focused exclusively on the \emph{r/incels} subreddit, which has historically been central to the development of incel ideologies on Reddit, but does not represent the full range of incel-related communities. Third, our operationalization of pre- vs.~post-engagement is simply based on the first active interaction with the incel community, whether in a comment or a post. This approach does not take into consideration passive participation, such as lurking or content consumption, which may also influence behavioral and psychological changes. Finally, our analysis of engagement in shared threads does not capture direct conversations among users, therefore provides a necessary but not a sufficient condition for social influence. 

Future research could address some of these limitations by analyzing data from multiple platforms and a wider range of subreddits or online communities to capture a more comprehensive view of radicalization processes. In addition, studying passive users, or lurkers, could shed light on the effects of exposure to these communities even in the absence of active participation. 
Scarce data availability from social media platforms, however, is posing increasing challenges to this research.